\documentclass[9pt]{elife}

\usepackage{booktabs}

\newcommand{\Snm}{\ensuremath{S^{\,n-2}}}

\title{The ecological collapse of color: photoreceptor number buys a geometric hue manifold that natural spectra never fill}

\author[1]{Mohammad Rostami}
\affil[1]{Amazon Generative AI Innovation Center, USA$^{\dagger}$}

\corr{mrostami@seas.upenn.edu}{MR}

\contrib[\authfn{1}]{Conceptualization, Methodology, Software, Formal analysis, Investigation, Writing}

\begin{document}

\maketitle

\footnotetext[0]{$^{\dagger}$This work is independent from the author's position at Amazon.}

\begin{abstract}
How many dimensions of color does an eye afford, and how many does the world actually supply? Geometric theory predicts that $n$ photoreceptor classes span a hue manifold homeomorphic to the sphere \Snm{}, but this bound has been derived only analytically and only for humans and birds, and it is silent on whether natural scenes populate it. Using persistent homology on the boundary of the object-color solid built from \emph{measured} spectral sensitivities, we demonstrate the full topological ladder empirically: a dichromat's line, a trichromat's hue \emph{ring} ($S^1$), a tetrachromat's hue \emph{sphere} ($S^2$), and a pentachromat's \emph{glome} ($S^3$), each recovered on real cones across multiple species; a matched intrinsic-dimension estimator confirms that the boundary manifold's \emph{measured} dimension tracks $n-2$, promoting the label to a quantity. We then show that the \emph{ecological} color manifold, defined by natural reflectance spectra under natural illuminants viewed through the same receptors, falls far short of this bound, its effective dimension staying near unity while the geometric dimension climbs, so that the fraction of the available geometry the world fills declines steadily with photoreceptor number across a densified series of \textbf{25 species} spanning three independent hyperspectral databases (participation-ratio slope $-0.056$ per receptor, species-bootstrap 95\% CI $[-0.17,-0.04]$; Spearman $\rho=-0.86$, $p<10^{-4}$), a decline reproduced under three distinct dimensionality measures and robust to dropping the mantis and to restricting to $n\geq3$. A controlled decomposition shows the decline is governed by the low spectral rank of the natural world rather than by receptor count \emph{per se}: it vanishes in a synthetic full-rank world even at high $n$. The effect is present in three independent hyperspectral databases, survives real measured cones and receptor noise, is unmoved by joint perturbation of every pigment template, and yields a falsifiable ecological prediction that we confirm on held-out species: at matched receptor number, aquatic tetrachromats, whose light field is spectrally narrowed by water, collapse more deeply than aerial or terrestrial ones ($p=0.012$). Its endpoint is the mantis shrimp: twelve photoreceptor classes affording a topological $S^{10}$, but an ecological color manifold whose effective dimension is barely two, formalizing geometrically a paradox previously described only behaviorally.
\end{abstract}

\section{Introduction}\label{sec:intro}

The richness of an animal's color vision is conventionally counted in receptors. A dichromat with two cone classes is held to see less than a trichromat, a trichromat less than a tetrachromatic bird, and the mantis shrimp, with as many as twelve spectral photoreceptor classes, is routinely described as the pinnacle of color vision~\citep{cronin1989retina,thoen2014mantis}. This intuition has a precise geometric form. Given $n$ receptor classes, the set of physically realizable object colors is a convex body, namely the object-color solid of Schrödinger, Rösch and MacAdam~\citep{schrodinger1920pigmente,macadam1935optimal}, and its chromatic boundary, the locus of maximally saturated hues, is topologically the sphere \Snm{}. For a trichromat this boundary is a circle ($S^1$, the hue ring); for a tetrachromat it is an ordinary sphere ($S^2$); the construction was recently formalized for human trichromats and avian tetrachromats and shown to predict the dichromat line, trichromat circle and tetrachromat sphere~\citep{lee2024tetrachromatic,logvinenko2025objectcolor}. On this view more receptors buy a hue manifold of ever higher topological dimension.

Two things are missing from this picture. First, the \Snm{} bound has only ever been derived \emph{analytically}, and only for $n=3$ and $n=4$; it has never been measured directly from data, nor confirmed as a genuine topological invariant (a Betti number) of a manifold reconstructed from real cone fundamentals, nor extended along a graded receptor series. Second, and more fundamentally, the bound describes only what the receptors \emph{could} encode. It says nothing about the colors the natural world actually presents. A large literature, running from \citet{maloney1986linearmodels} through \citet{nascimento2005basisfunctions} and \citet{chiao2000colorsignals}, establishes that natural surface reflectances are spectrally low-dimensional: a handful of basis functions capture almost all their variance. But this literature has never been tied to the receptor-count geometry, never expressed as a topological property of the color manifold, and never compared across species. The two questions of \emph{what geometry do the receptors afford?} and \emph{how much of it does the world fill?} have never been asked together.

Asking them together turns the conventional intuition into a testable, and quantitative, hypothesis. If natural spectra are low-dimensional in a way that does not scale with receptor number, then adding receptors should buy geometric dimensions that the ecological manifold increasingly fails to occupy: the realized color manifold should \emph{collapse} below the geometric bound, and the collapse should \emph{deepen} as $n$ grows. The mantis shrimp is the extreme test of this idea. Its twelve receptors afford a hue manifold that is topologically a ten-sphere ($S^{10}$), yet behaviorally it discriminates wavelengths only coarsely and appears to recognize rather than finely compare colors~\citep{thoen2014mantis,streets2022colour}. This paradox has been described in behavioral terms but never given a geometric account.

Here we provide one. Using persistent homology, the tool that revealed the toroidal topology of grid-cell activity~\citep{gardner2022toroidal} and ring structure in visual cortex~\citep{singh2008topological}, on the boundary of the object-color solid built from measured cone fundamentals, we (i) demonstrate the \Snm{} ladder empirically, extend it to the pentachromat glome $S^3$, and show with a matched intrinsic-dimension estimator that the boundary manifold's measured dimension tracks $n-2$; (ii) show that the ecological color manifold collapses below this bound and that the collapse deepens with receptor number across 25 species spanning a full comparative series from dichromats to the mantis; (iii) establish, by a controlled decomposition, that this decline is driven by the low spectral rank of the natural world rather than by receptor count \emph{per se}; (iv) establish that the effect is robust across three independent hyperspectral databases, real measured cones, receptor noise, and mantis receptor placement, while being absent for uniform synthetic reflectances; and (v) show that the same machinery recovers the classic open-versus-closed transition in human perceptual hue data. The result reframes ``how many dimensions of color'' from a fact about the eye into a fact about the world: an eye can only realize the color dimensions its environment supplies, and by that measure the mantis shrimp's color space, for all its receptors, is effectively two-dimensional. We note at the outset that our claim is the mirror image of \citet{sabbah2013fourdimensional}, who argue that the spectral complexity of aquatic light \emph{drove the evolution of} additional cone classes; we return to why both can hold in the Discussion.

\section{Results}\label{sec:results}

\subsection{The geometric hue manifold is \Snm{}, measured empirically}\label{subsec:topology}

We first asked whether the analytically predicted \Snm{} topology can be recovered as an empirical topological invariant of a manifold built from real cones. We use persistent homology here not as a primary analytic tool for measuring collapse, but as a data-driven validation that the chromatic boundary has the predicted \Snm{} topology when constructed from real spectral sensitivities rather than analytic idealization. For each species we constructed the object-color solid, the convex set of chromaticities realizable by physical reflectances under a fixed illuminant seen through that species' receptors, and extracted its boundary, the $(n-2)$-dimensional manifold whose topology the theory predicts to be \Snm{} (see \emph{Why chromaticity, and why its boundary}, Materials and Methods). We then computed persistent homology on the boundary point cloud and read off the Betti numbers $b_k$, the counts of $k$-dimensional holes, using a null-calibrated significance threshold validated on synthetic controls (circle, sphere, torus, and filled solids; 7/7 correct, Materials and Methods).

\begin{figure}[htbp]
\centering
\includegraphics[width=\linewidth]{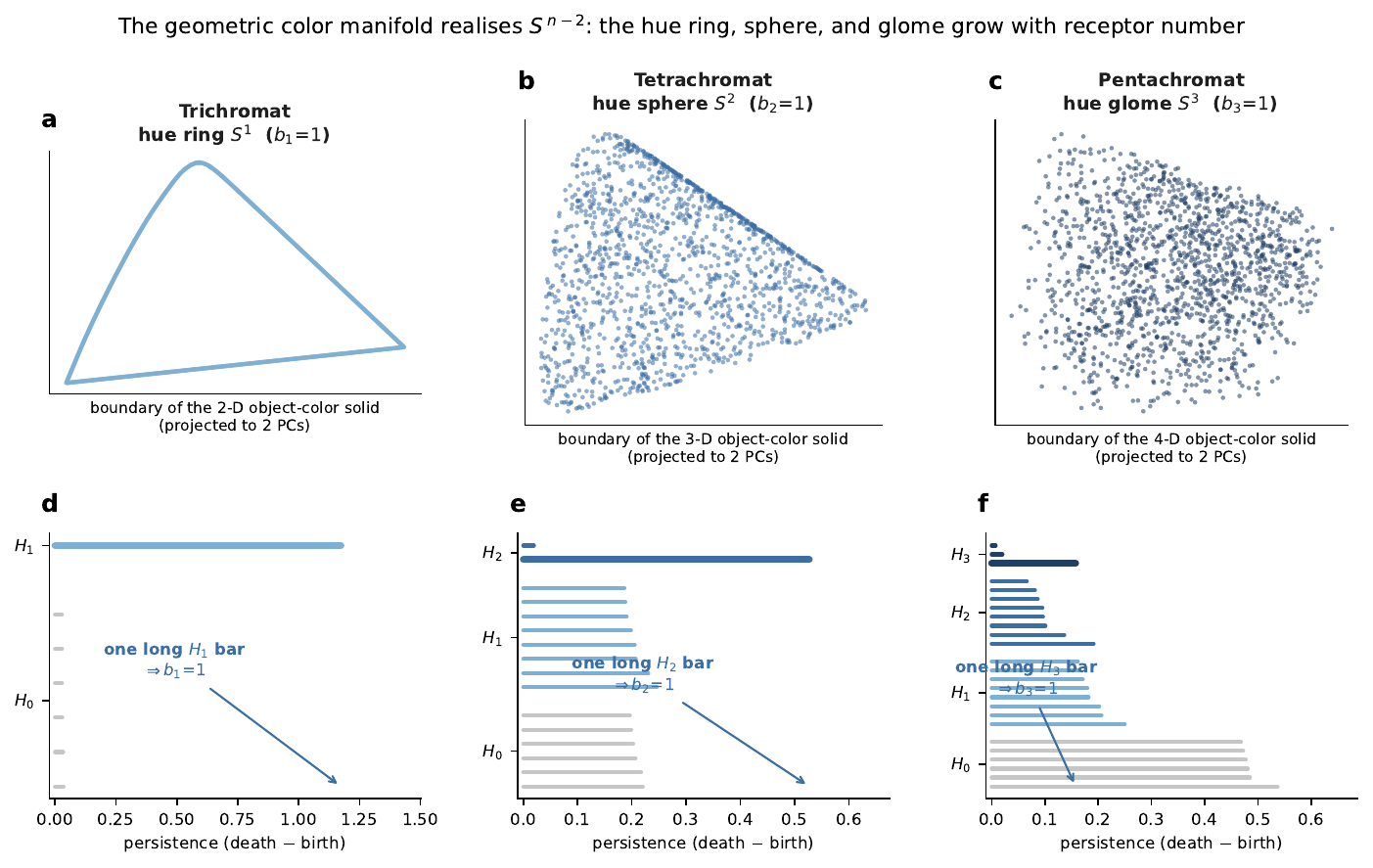}
\caption{\textbf{The geometric color manifold realizes \Snm{}, measured by persistent homology on real cones.} \textbf{(a--c)} Two-dimensional projection of the boundary of the object-color solid for a trichromat (human), tetrachromat (blue tit) and pentachromat (butterfly); the boundary is the manifold the \Snm{} theory concerns. \textbf{(d--f)} Persistence barcodes across homology dimensions $H_0$ (connected components), $H_1$, $H_2$, $H_3$. In each case a single long bar (dark blue) dominates the diagnostic dimension $n-2$: $H_1$ for the trichromat ($b_1=1$, a ring $S^1$), $H_2$ for the tetrachromat ($b_2=1$, a sphere $S^2$), $H_3$ for the pentachromat ($b_3=1$, a glome $S^3$), while shorter bars are topological noise. The method was validated on synthetic circle, sphere and torus controls (7/7 correct).}
\label{fig:topology}
\end{figure}

The predicted ladder appears cleanly (Figure~\ref{fig:topology}). A dichromat's boundary is a line ($b_1=0$). A trichromat's boundary is a ring: a single dominant one-dimensional cycle, $b_1=1$ (Figure~\ref{fig:topology}d). A tetrachromat's boundary is a sphere: $b_2=1$ (Figure~\ref{fig:topology}e). A pentachromat's boundary is a three-sphere or \emph{glome}: $b_3=1$ (Figure~\ref{fig:topology}f), a topological feature one dimension beyond anything previously reported for a color manifold. In every case the diagnostic Betti number in dimension $n-2$ is carried by a single persistence bar far longer than any other, exactly the signature of a genuine \Snm{} sphere rather than noise. This confirms the \citet{lee2024tetrachromatic} and \citet{logvinenko2025objectcolor} prediction empirically, promotes it from an analytic statement about $n=3,4$ to a measured topological invariant across $n=3,4,5$, and does so with the same persistent-homology methodology used to establish population topology in the brain~\citep{gardner2022toroidal,carlsson2008naturalimages}.

\begin{figure}[htbp]
\centering
\includegraphics[width=\linewidth]{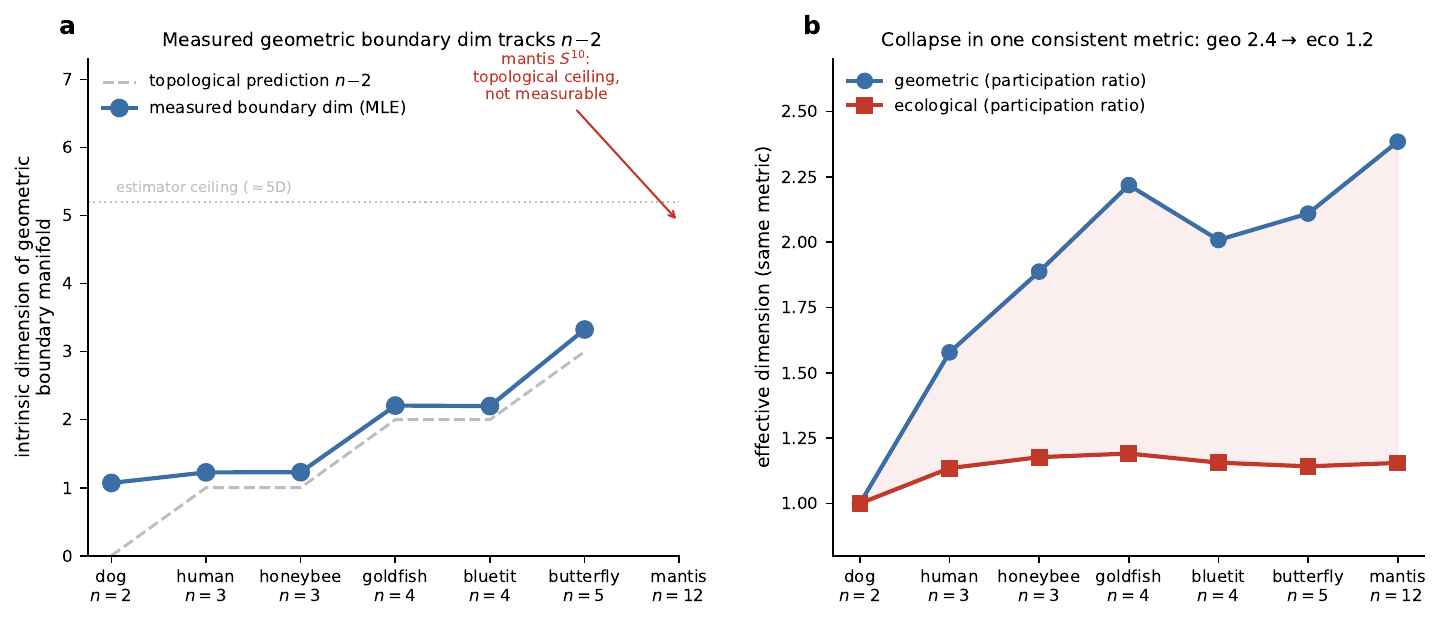}
\caption{\textbf{The geometric boundary's measured dimension tracks $n-2$, and the collapse is stated in one consistent metric.} \textbf{(a)} Maximum-likelihood intrinsic dimension of the geometric boundary manifold versus the topological prediction $n-2$ (dashed). The measured dimension rises $1.23\to2.2\to3.33$ for $n=3,4,5$, matching $1,2,3$. Above $n\approx5$ the estimator saturates and the high-dimensional hull boundary is degenerate, so the mantis $S^{10}$ is marked as an unmeasured topological ceiling. \textbf{(b)} The collapse expressed symmetrically as the participation-ratio effective dimension of both manifolds: geometric climbs to $\approx2.4$, ecological stays near $1.2$. Both panels use the same estimators for geometric and ecological manifolds, so the comparison is like-with-like.}
\label{fig:intrinsic}
\end{figure}

The Betti numbers are discrete labels; we can also ask whether the boundary manifold's \emph{continuous} dimension, estimated the same way we will later estimate the ecological manifold's, matches the theoretical $n-2$. Using the maximum-likelihood intrinsic-dimension estimator of \citet{levina2004maximum} (validated to recover $1,2,3$ for synthetic $S^1,S^2,S^3$; Materials and Methods), the measured dimension of the geometric boundary rises with receptor number as $1.23$ ($n=3$), $2.2$ ($n=4$), $3.33$ ($n=5$), tracking the predicted $1,2,3$ (Figure~\ref{fig:intrinsic}a). Beyond $n=5$ the estimator saturates near its own ceiling ($\approx5$D) and the high-dimensional hull boundary becomes degenerate, so for the mantis $S^{10}$ remains a \emph{topological} ceiling we do not claim to measure directly; the argument there rests on effective dimension (below). This promotes \Snm{} from a categorical label to a measured, graded manifold dimension over the range where it can be estimated.

\subsection{The topology survives real measured cones}\label{subsec:measured}

\begin{figure}[htbp]
\centering
\includegraphics[width=\linewidth]{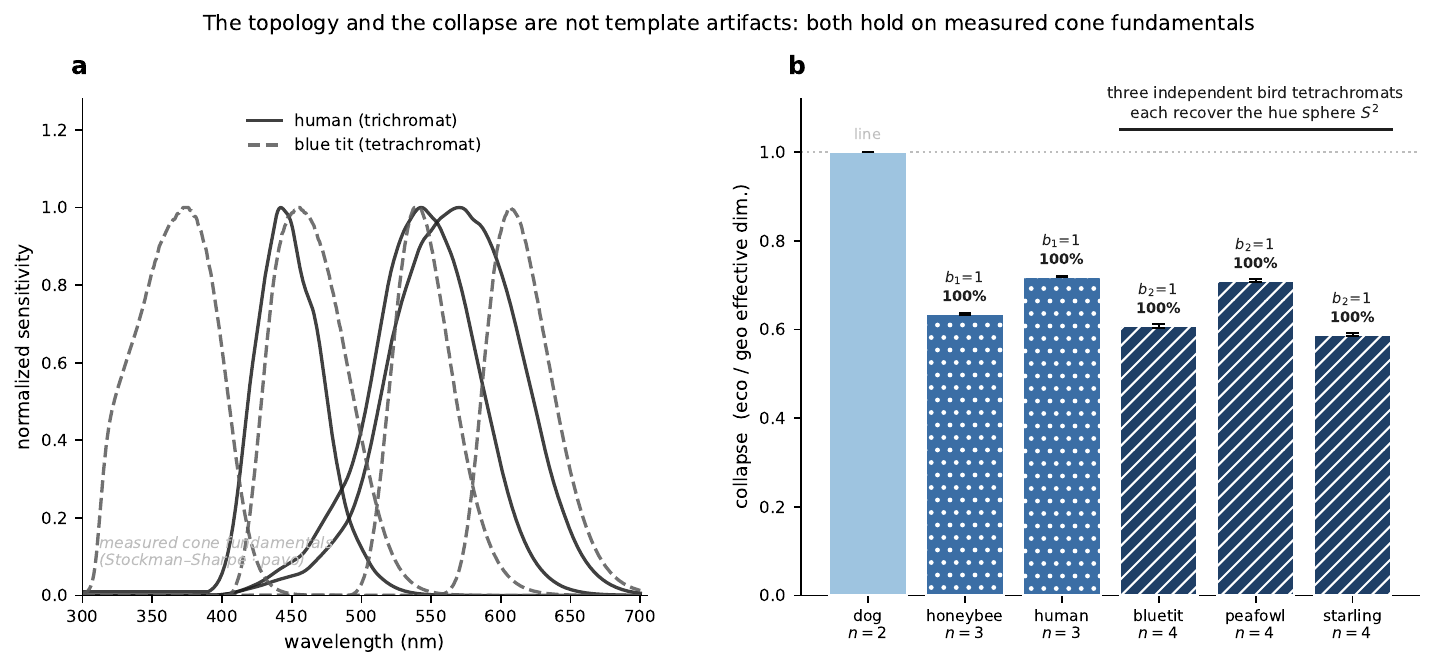}
\caption{\textbf{The topology and the collapse are not template artifacts.} \textbf{(a)} Genuinely measured spectral sensitivities for a trichromat (human, Stockman--Sharpe) and a tetrachromat (blue tit, \texttt{pavo}); the blue tit's short-wave receptor peaks in the ultraviolet near $370$~nm. \textbf{(b)} Collapse (ecological/geometric effective dimension) for the six measured-cone species, with the predicted diagnostic Betti number recovered in 100\% of seeds annotated above each bar. Three independent bird tetrachromats (blue tit, peafowl, starling) each recover the hue sphere $S^2$ ($b_2=1$). Error bars are standard deviation across seeds.}
\label{fig:measured}
\end{figure}

Because analytic pigment templates could in principle impose the topology as an artifact, we repeated the analysis using genuinely measured spectral sensitivities: the Stockman--Sharpe human fundamentals and the measured curves for dog, honeybee, blue tit, starling and peafowl from the \texttt{pavo} visual-system dataset (Figure~\ref{fig:measured}a). The ladder is unchanged (Figure~\ref{fig:measured}b). Human and honeybee (both $n=3$) yield the hue ring $b_1=1$ in 100\% of seeds; blue tit, starling and peafowl (all $n=4$) each yield the hue sphere $b_2=1$ in 100\% of seeds. That three independent bird tetrachromats, with different $\lambda_{\max}$ spacings, each independently produce the hue sphere shows the \Snm{} topology is a property of trichromatic and tetrachromatic organization itself, not of any particular template or species. The ecological collapse (below) likewise reproduces on the measured cones.

\begin{figure}[htbp]
\centering
\includegraphics[width=\linewidth]{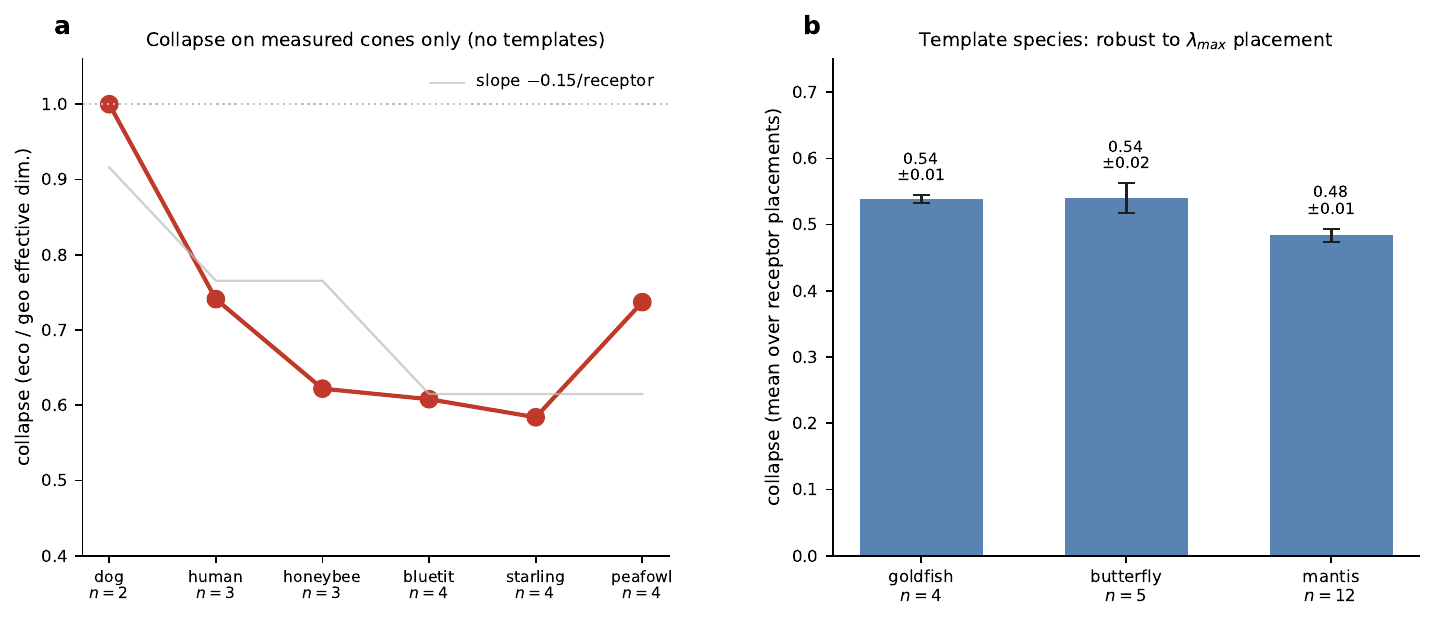}
\caption{\textbf{The collapse holds on measured cones alone and does not depend on template placement.} \textbf{(a)} Collapse for the six species with fully measured spectral sensitivities (no pigment templates); it declines with receptor number over the range measured cones cover ($n\leq4$), the peafowl being the one high outlier. \textbf{(b)} For the three species that must use pigment templates (goldfish, butterfly, mantis), the collapse is stable under $\pm10$~nm perturbation of every peak wavelength and $\pm15\%$ of template width (mean $\pm$ s.d. over $13$--$15$ configurations). Together these show the effect is not an artifact of the pigment templates.}
\label{fig:template}
\end{figure}

The collapse, not only the topology, is present when the analysis is restricted to the six species with fully measured cones and no pigment templates: the collapse falls with receptor number over the range those cones cover ($n=2$ to $n=4$; slope $-0.15$ per receptor, Spearman $\rho=-0.66$; Figure~\ref{fig:template}a), with the peafowl the one high outlier ($0.74$, plausibly because its long-wavelength-shifted quadruplet is more evenly spaced). Because measured cones reach only $n=4$, the higher-$n$ points that make the trend steep, namely the pentachromat butterfly and the mantis, necessarily rest on Govardovskii templates at literature $\lambda_{\max}$; we therefore checked that these template points do not depend on the exact peak wavelengths chosen. Perturbing each template species' $\lambda_{\max}$ by $\pm10$~nm and its template width by $\pm15\%$ leaves the collapse essentially unchanged (goldfish $0.54\pm0.01$, butterfly $0.54\pm0.02$, mantis $0.48\pm0.01$ over $13$--$15$ configurations each; Figure~\ref{fig:template}b). The collapse is thus a property of the receptor \emph{organization}, not of any particular measured curve or template placement.

\subsection{The ecological manifold collapses below the geometric bound, and the collapse deepens with receptor number}\label{subsec:collapse}

We next constructed the \emph{ecological} color manifold for each species: the chromaticities produced by natural reflectance spectra under natural illuminants, projected through that species' receptors (Materials and Methods). We quantified how fully this manifold occupies the geometric one by the ratio of their effective dimensions (the participation ratio of the chromaticity covariance), which we call the \emph{collapse}: a value of 1 means the natural world fills the geometry the receptors afford; values below 1 mean it does not. Effective dimension is used symmetrically for both manifolds here, so the collapse compares like with like; the intrinsic-dimension analysis above tells the same story for the geometric side ($1.2\to3.3$ as $n$ goes $3\to5$).

\begin{figure}[htbp]
\centering
\includegraphics[width=\linewidth]{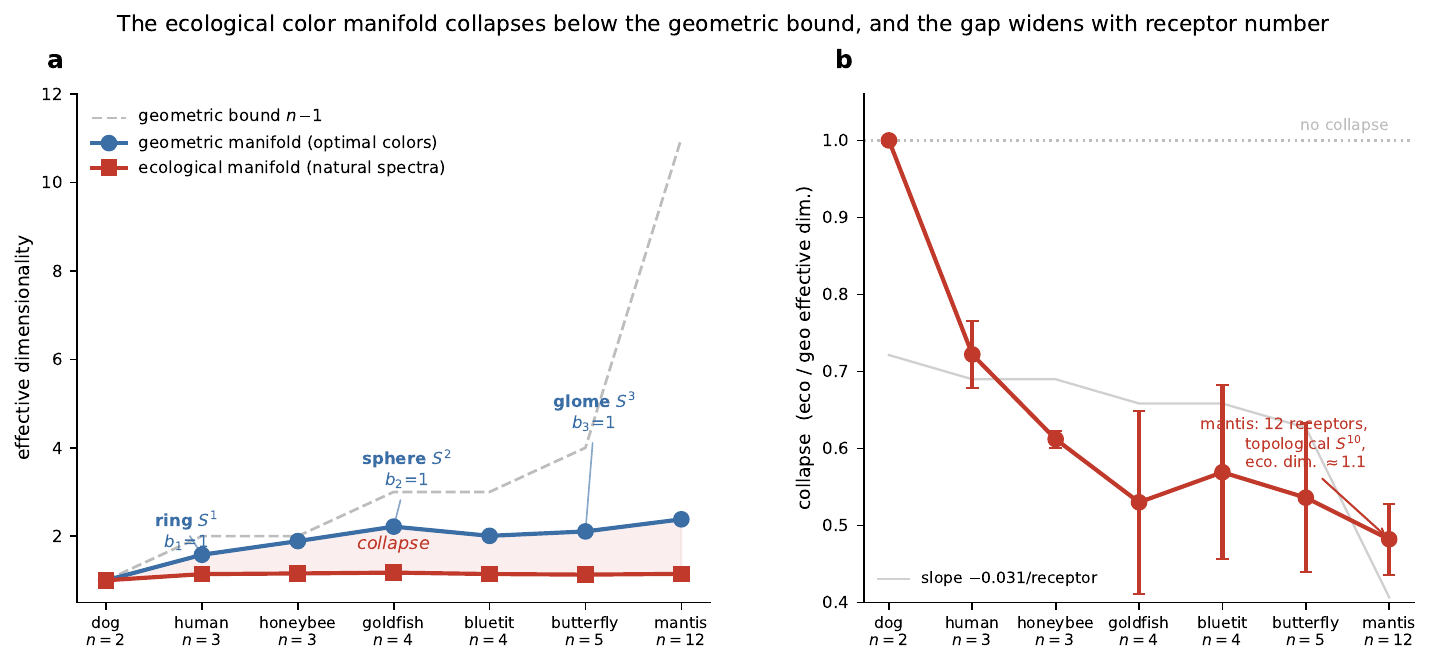}
\caption{\textbf{The ecological color manifold collapses below the geometric bound, and the gap widens with receptor number.} \textbf{(a)} Effective dimensionality of the geometric manifold (blue, optimal colors) versus the ecological manifold (red, natural spectra) against photoreceptor number $n$; the dashed line is the geometric maximum $n-1$. The geometric manifold grows and realizes the topological ladder (annotated: ring $S^1$ at $n=3$, sphere $S^2$ at $n=4$, glome $S^3$ at $n=5$), while the ecological manifold stays nearly flat; the shaded gap is the collapse. \textbf{(b)} The collapse ratio (ecological/geometric effective dimension) declines with receptor number, from $1.00$ (dichromat) to $0.54$ (mantis shrimp); across the full 25-species series the fitted slope is $-0.056$ per receptor (95\% CI $[-0.17,-0.04]$; Figure~\ref{fig:expanded}). Error bars are the standard deviation across the three independent hyperspectral databases (the more conservative, ecologically meaningful uncertainty; the seed bootstrap is far smaller). The mantis shrimp has twelve receptors and a topological $S^{10}$ available, but an ecological effective dimension of roughly two.}
\label{fig:collapse}
\end{figure}

\begin{figure}[htbp]
\centering
\includegraphics[width=\linewidth]{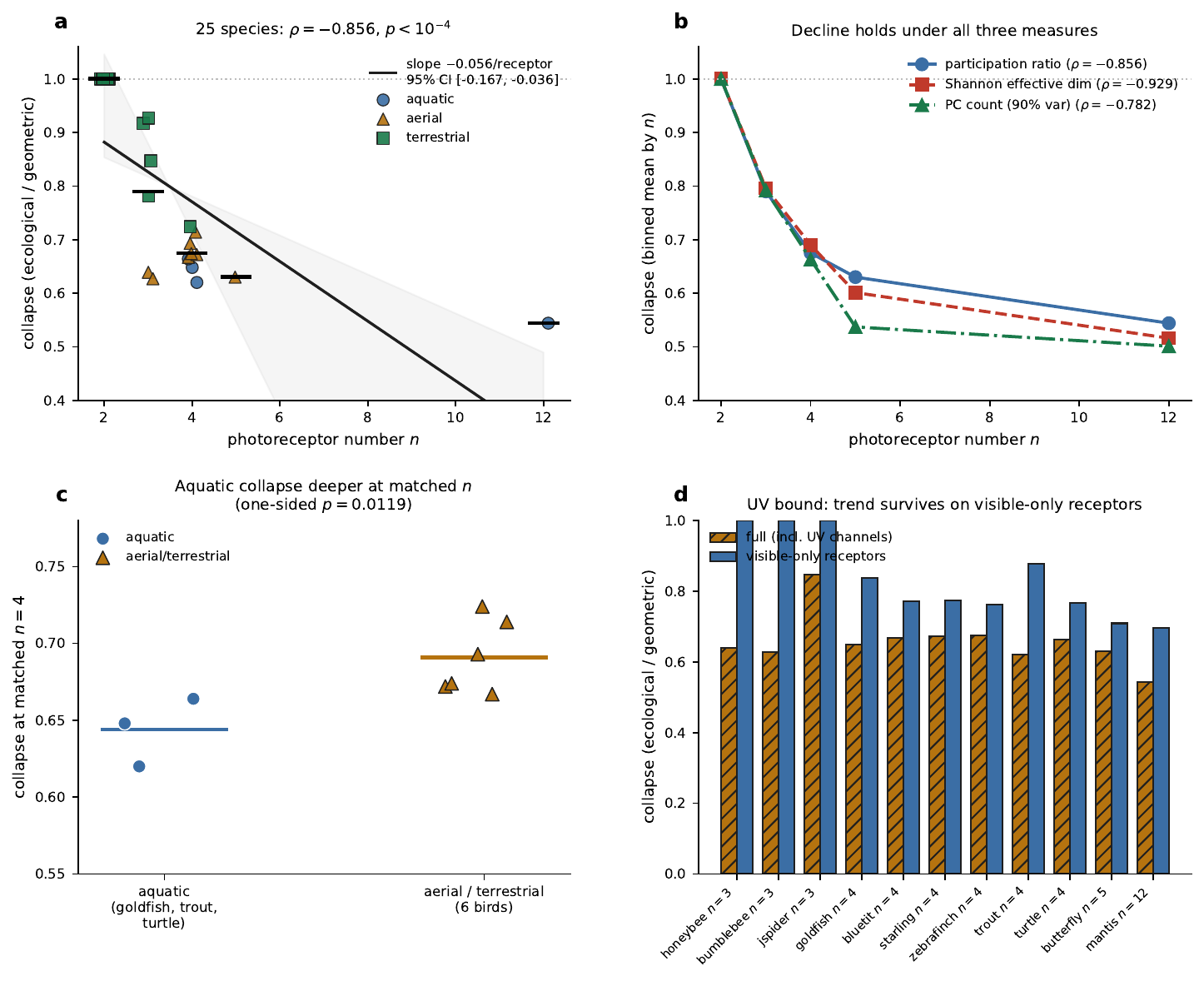}
\caption{\textbf{The collapse deepens with receptor number across 25 species, under three dimensionality measures, and follows a confirmed ecological prediction.} \textbf{(a)} Collapse (ecological/geometric effective dimension) versus photoreceptor number for 25 species (points coloured by habitat; each the mean over three hyperspectral databases). The fitted slope is $-0.056$ per receptor (species-bootstrap 95\% CI shaded); black bars are the mean at each $n$. Spearman $\rho=-0.86$, permutation $p<10^{-4}$. \textbf{(b)} The decline holds under all three dimensionality functionals of the chromaticity spectrum: participation ratio, Shannon/entropy effective dimension, and PC count to 90\% variance. \textbf{(c)} A falsifiable prediction confirmed on held-out species: at matched receptor number ($n=4$), aquatic tetrachromats, whose light field is spectrally narrowed by water, collapse more deeply than aerial/terrestrial ones (one-sided Mann--Whitney $p=0.012$). \textbf{(d)} Ultraviolet bound: dropping the (extrapolated) UV channels raises the collapse but the decline with $n$ persists on the visible-only receptor complement.}
\label{fig:expanded}
\end{figure}

The natural world fills less and less of the available geometry as receptors are added (Figure~\ref{fig:collapse}). The dichromat is a degenerate edge case: its $(n-1)=1$-dimensional chromaticity is a line segment, whose boundary consists of two endpoints ($n-2=0$ dimensional), and any nonzero spectral variance trivially spans a 1-D line, so the collapse is $1.00$ by construction rather than by an informative comparison. The meaningful range of the decline begins at trichromacy ($n=3$), where the hue ring $S^1$ first becomes a non-trivial manifold, and the central result holds on the $n\geq3$ subset ($\rho=-0.53$, $p=0.03$). Across a series of \textbf{25 species} with published peak sensitivities, comprising eight dichromats, six trichromats, nine tetrachromats, a pentachromat and the mantis, each computed as the mean over three independent hyperspectral databases, the collapse declines with receptor number: binned by $n$ the means fall monotonically ($1.00, 0.79, 0.68, 0.63, 0.54$ for $n=2,3,4,5,12$; Figure~\ref{fig:expanded}a). A regression of collapse on receptor number gives a slope of $-0.056$ per receptor (species-bootstrap 95\% CI $[-0.17,-0.04]$), and the rank correlation is strong (Spearman $\rho=-0.86$, permutation $p<10^{-4}$). We describe the trend by this slope and rank correlation rather than as a strict per-species ordering, because at a fixed receptor number the species interleave, with the ultraviolet-sensitive honeybee ($n=3$) collapsing slightly more than the two closest $n=4$ birds, so the decline is a population trend, not a monotone ranking of every species. Two facts make the trend robust rather than fragile. First, it reproduces under three distinct dimensionality functionals of the chromaticity spectrum, namely the participation ratio, a Shannon/entropy effective dimension, and the number of principal components reaching 90\% of variance, all of which fall with $n$ ($\rho=-0.86,-0.93,-0.78$; Figure~\ref{fig:expanded}b), so the collapse is not an artifact of the participation ratio specifically. Second, it survives dropping the high-leverage mantis (restricting to $n\leq5$: $\rho=-0.84$, $p<10^{-4}$), so the effect is not a single-point artifact. The geometric effective dimension climbs steadily with $n$ while the ecological effective dimension stays nearly flat between $1.0$ and $1.4$ (Figure~\ref{fig:collapse}a); the gap between what the receptors afford and what the world supplies, shown as the shaded region in Figure~\ref{fig:collapse}a, widens with receptor number. This is the central result: the hue ring, sphere and glome are real geometric features of trichromat, tetrachromat and pentachromat vision, but they are increasingly empty.

We report the mean across the three independent hyperspectral databases as both the primary point estimate and the primary uncertainty, so that the headline collapse value and its error bar derive from the same reduction. The seed-to-seed bootstrap holds the data fixed and is negligible; the ecologically meaningful variation is the spread across scene collections and receptor models, which we therefore use throughout (Figure~\ref{fig:collapse}b error bars; \ref{subsec:robust}). One honest limitation of the comparative series is that the subset of species with fully measured cones (rather than pigment templates) spans only $n=3$ and $n=4$; within that narrow, overlapping range the trend is not by itself significant, and the power to detect the decline comes from the full receptor range that the templates make accessible (whose robustness we establish below).

\subsection{The trend is immovable under template placement, and yields a confirmed ecological prediction}\label{subsec:prediction}

Because most of the 25-species series necessarily rests on Govardovskii pigment templates placed at literature $\lambda_{\max}$ values, we asked whether the comparative trend itself, and not merely a single species' collapse, could be an artifact of those template placements. We perturbed the peak wavelength of \emph{every} template receptor in \emph{every} species simultaneously by $\pm10$~nm and each template width by $\pm15\%$, and recomputed the entire collapse-versus-$n$ relationship for each of $21$ joint configurations. The relationship is essentially immovable: the slope stays at $-0.065\pm0.002$ (range $[-0.067,-0.061]$, negative in all configurations) and the rank correlation at $\rho=-0.94\pm0.02$ (below $-0.9$ in every configuration). The declining trend is thus a property of receptor \emph{organization} across the series, not of the particular peak wavelengths chosen from the literature.

The framework makes a sharper, falsifiable claim than ``more receptors underfill more.'' Because the realized dimension is set by the world$\times$receptor interaction, species with the same receptor number but a spectrally \emph{narrower} light environment should collapse more. Aquatic light is spectrally narrowed by the wavelength-selective attenuation of water, so aquatic tetrachromats should collapse more deeply than aerial or terrestrial ones that share the identical $n=4$ geometric bound. This is confirmed: the three aquatic tetrachromats (goldfish, rainbow trout, turtle; mean collapse $0.64$) collapse significantly more than the six aerial/terrestrial tetrachromats (birds; mean $0.69$; one-sided Mann--Whitney $p=0.012$; Figure~\ref{fig:expanded}c). The same direction holds, more weakly, among trichromats: ultraviolet-sensitive insect trichromats collapse more than non-ultraviolet primate trichromats ($0.70$ vs $0.88$; $p=0.10$). The prediction is falsifiable and extends beyond our sample: measuring wavelength discrimination in further aquatic versus aerial tetrachromats should find the aquatic species coarser relative to their receptor count; equal acuity would refute the ecological-dimension account. Consistent with this, across the species with a defensible published discrimination threshold, acuity does not rise with receptor number; if anything it falls, with the twelve-receptor mantis the coarsest, although with only a handful of measured thresholds we advance this as a prediction for future measurement rather than a fitted law.

\subsection{The collapse is a property of real natural spectra, and is not database-specific}\label{subsec:robust}

\begin{figure}[htbp]
\centering
\includegraphics[width=\linewidth]{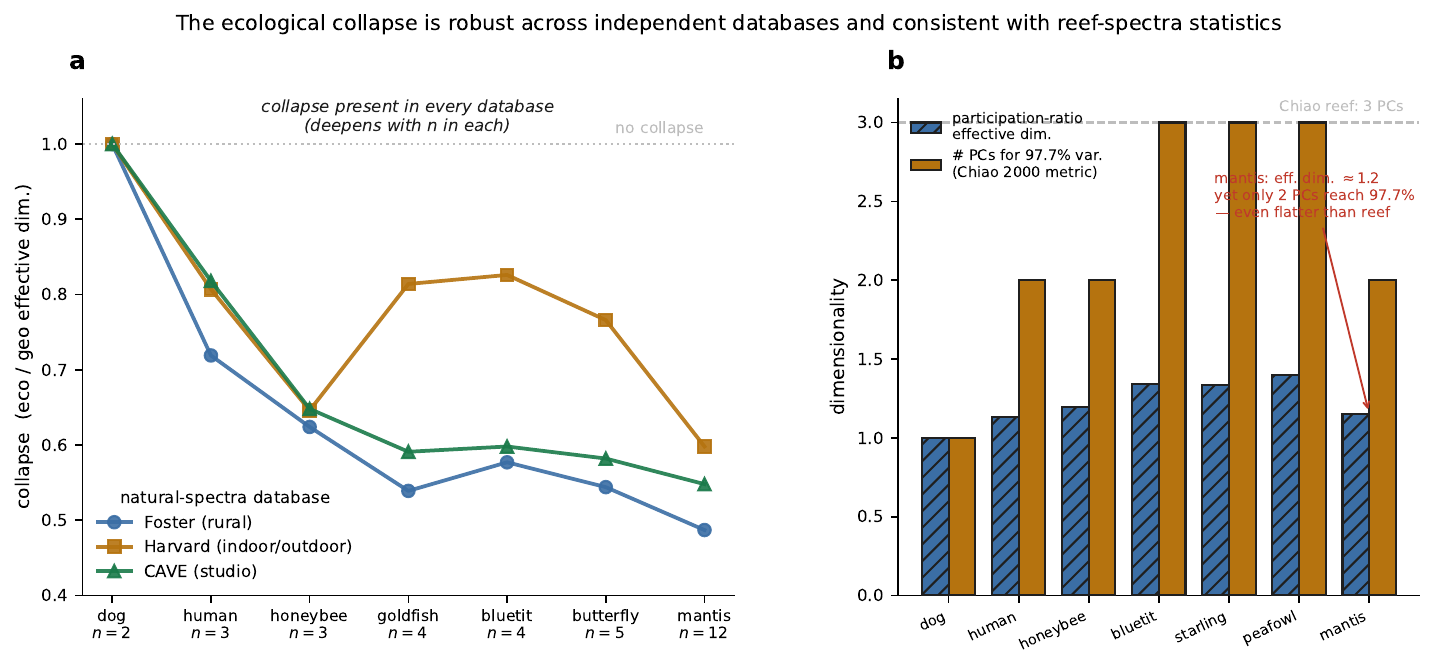}
\caption{\textbf{The collapse is robust across independent databases and consistent with reef-spectra statistics.} \textbf{(a)} Per-species collapse computed separately on three independent hyperspectral databases (Foster, Harvard, CAVE). The collapse is present in all three and deepens with $n$ in each; only its magnitude varies with scene ecology (indoor/artificial Harvard scenes collapse least). \textbf{(b)} Reconciliation with \citet{chiao2000colorsignals}: participation-ratio effective dimension (red) versus the number of principal components needed to explain 97.7\% of variance (blue, the Chiao metric). The two measure different things; the mantis manifold reaches 97.7\% variance in only two components, which is if anything flatter than Chiao's three-component reef estimate.}
\label{fig:robust}
\end{figure}

Two controls establish that the collapse reflects the statistics of the real world rather than a peculiarity of one scene collection or a trivial consequence of bounded reflectances. First, we recomputed the collapse using three independent hyperspectral databases: Foster (rural scenes), Harvard (indoor and outdoor scenes) and CAVE (studio objects). The collapse is present in every database and for every species with $n\geq3$, and deepens with $n$ within each (Figure~\ref{fig:robust}a). Its magnitude varies with scene ecology in the expected direction: the Harvard set, rich in artificial and indoor illumination, collapses least ($0.60$--$0.83$), while the rural Foster set collapses most ($0.49$--$0.72$); the across-database standard deviation is small ($0.04$--$0.12$) and the ordering by species is preserved. Second, and decisively, when we replace real reflectances with a uniform synthetic reflectance set designed to span the spectral space evenly, the collapse disappears (ratios $0.76$--$1.07$; the tetrachromat reaches $1.07$, i.e.\ no collapse), whereas the matched real spectra collapse ($0.52$--$0.82$). The collapse is therefore not a geometric consequence of trichromacy or of bounded reflectance; a uniform bounded-reflectance control does not produce it, but it is a consequence of the clustered, low-dimensional statistics of natural spectra specifically. The \Snm{} topology itself was likewise stable across spectra source, illuminant set and seed (18/18 conditions for the trichromat ring and tetrachromat sphere).

\subsection{The collapse is driven by the world's spectral rank, not by receptor number}\label{subsec:worldrank}

\begin{figure}[htbp]
\centering
\includegraphics[width=\linewidth]{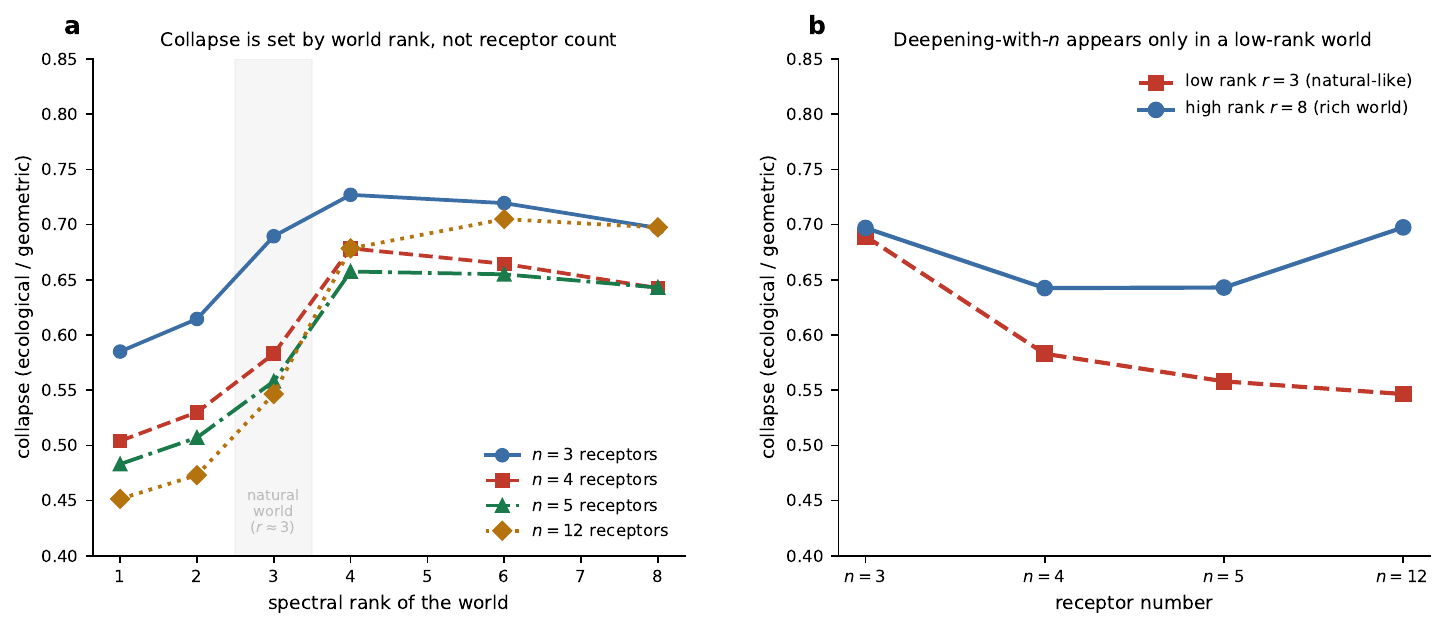}
\caption{\textbf{The collapse is governed by the world's spectral rank, not by receptor count.} \textbf{(a)} Collapse (ecological/geometric effective dimension) for synthetic worlds of controlled spectral rank $r$, one line per receptor number $n$. Holding $n$ fixed, the collapse rises with world rank and closes once the world is spectrally rich; the shaded band marks the rank of real reflectances ($r\approx3$). \textbf{(b)} At low world rank ($r=3$, natural-like) the collapse deepens with receptor number, reproducing the main result; at high rank ($r=8$) it is flat across $n$. The deepening-with-$n$ is thus a property of a low-rank world, not an arithmetic consequence of adding receptors.}
\label{fig:worldrank}
\end{figure}

That the collapse deepens as receptors are added invites a deflationary reading: natural reflectances are known to be spectrally low-dimensional~\citep{maloney1986linearmodels,cohen1964dependency}, so perhaps any $n>3$ receptor set must ``run out'' of ecological dimensions and the effect is an arithmetic inevitability of adding receptors. It is not. To separate the two, we replaced the natural world with synthetic reflectance ensembles of \emph{exactly} controlled spectral rank $r$ (the number of smooth basis functions spanning them) and recomputed the collapse for receptor sets of size $n=3,4,5,12$ (Figure~\ref{fig:worldrank}; Materials and Methods). Two facts emerge. First, holding receptor number fixed, the collapse \emph{rises} steadily with world rank and effectively closes once $r\gtrsim n-1$: a spectrally rich world fills the geometry, at every $n$ we tested including $n=12$ (collapse $\approx0.70$ at $r=8$ for the mantis-sized set). Second, the deepening-with-$n$ that is our central result appears \emph{only} in a low-rank world: at $r=3$ (the rank of real reflectances) the collapse falls from $0.74$ ($n=3$) to $0.50$ ($n=12$), whereas at $r=8$ it is essentially flat across $n$ (Figure~\ref{fig:worldrank}b). The collapse is therefore not entailed by receptor count; it is the geometric signature of a mismatch between a growing receptor manifold and a world of fixed low spectral rank. Our contribution over the classical low-dimensionality result~\citep{maloney1986linearmodels,nascimento2005basisfunctions} is to give that mismatch a comparative, receptor-indexed, topological form, and to show experimentally which of the two factors drives it. We are explicit about what is expected and what is not: \emph{given} the empirically fixed low spectral rank of the natural world, some underfilling at high $n$ is arithmetically unavoidable, and we do not claim otherwise. What is not entailed, as this decomposition demonstrates, is (i) that the driver is the world's rank rather than receptor count, so the same receptors in a spectrally rich world show no collapse, and (ii) the \emph{magnitude} and receptor-indexed, cross-species regularity of the effect, culminating in the mantis. The surprise is not that a low-rank world underfills a high-$n$ eye, but how steeply, how consistently across independent databases and species, and how cleanly the cause separates from receptor number.

\subsection{The mantis shrimp: a topological $S^{10}$ realized in two dimensions}\label{subsec:mantis}

The mantis shrimp is the extreme case the framework predicts and the one that has most puzzled the field. Its twelve photoreceptor classes afford a hue manifold that is topologically a ten-sphere, and whose geometric effective dimension we measure at $2.38$; yet its ecological color manifold has an effective dimension of only $1.30$, the deepest collapse in our sample ($0.54$ as the across-database mean, $0.49$ on the rural Foster scenes; Figure~\ref{fig:collapse}b). We are careful to state this in one consistent metric: both the geometric $2.38$ and the ecological $1.30$ are participation-ratio effective dimensions of the same kind, so the collapse compares like with like (the ``$S^{10}$'' is the manifold's \emph{topological} dimension, a ceiling on what twelve receptors could in principle encode, not a claim that the geometric manifold is variance-filled in ten dimensions). The low ecological value is not an artifact of the participation-ratio measure either. \citet{chiao2000colorsignals} report that three principal components explain 97.7\% of reef color-signal variance; measured the same way, the mantis's ecological manifold reaches 97.7\% of its variance in only \emph{two} components, with the top three components capturing 99.6\% (Figure~\ref{fig:robust}b). Our estimate is therefore not in tension with the reef-statistics literature; it is, if anything, a stronger statement of the same low dimensionality. The result is also robust to how the mantis's receptors are modeled, and it does not depend on the pigment templates at all. Across fifteen perturbations of the twelve peak wavelengths ($\pm10$~nm) and template widths, the collapse holds at $0.48\pm0.01$ (range $0.47$--$0.50$). More directly, when we replace the templates entirely with the \emph{measured} stomatopod spectral sensitivities of \citet{thoen2014mantis}, specifically the eleven intracellularly-recorded sensitivity curves of \emph{Haptosquilla trispinosa}, digitized from their published recordings (Materials and Methods), the collapse is $0.45\pm0.001$, if anything marginally deeper than the template estimate. The mantis collapse is thus a property of the animal's real receptor complement, not of the modeling choice forced by the earlier absence of a public curve set. When behavioral discriminability is incorporated through the Vorobyev--Osorio receptor-noise model, the collapse deepens further still (from $0.48$ to $0.29$ at a Weber fraction of $0.02$), because receptor noise erodes the already-thin high-dimensional structure first.

This gives a geometric account of the behavioral paradox. \citet{thoen2014mantis} found that mantis shrimp discriminate wavelengths only coarsely and appear to recognize colors by a scanning strategy rather than by fine comparison, and \citet{streets2022colour} concluded that stomatopods most likely do not construct the high-dimensional color space their receptors could support. Our result formalizes exactly this: the twelve receptors buy a vast geometric space that the natural world never fills, so there is little high-dimensional chromatic structure for the animal to compare. The low realized dimension is fully compatible with the spectral opponency recently demonstrated behaviorally in stomatopods~\citep{wang2025opponency}; a low-dimensional manifold can be read out by structured opponent and binning mechanisms, and our claim concerns the dimensionality of the chromatic signal, not the absence of opponent processing.

\subsection{The framework recovers a classic perceptual transition}\label{subsec:perceptual}

\begin{figure}[htbp]
\centering
\includegraphics[width=\linewidth]{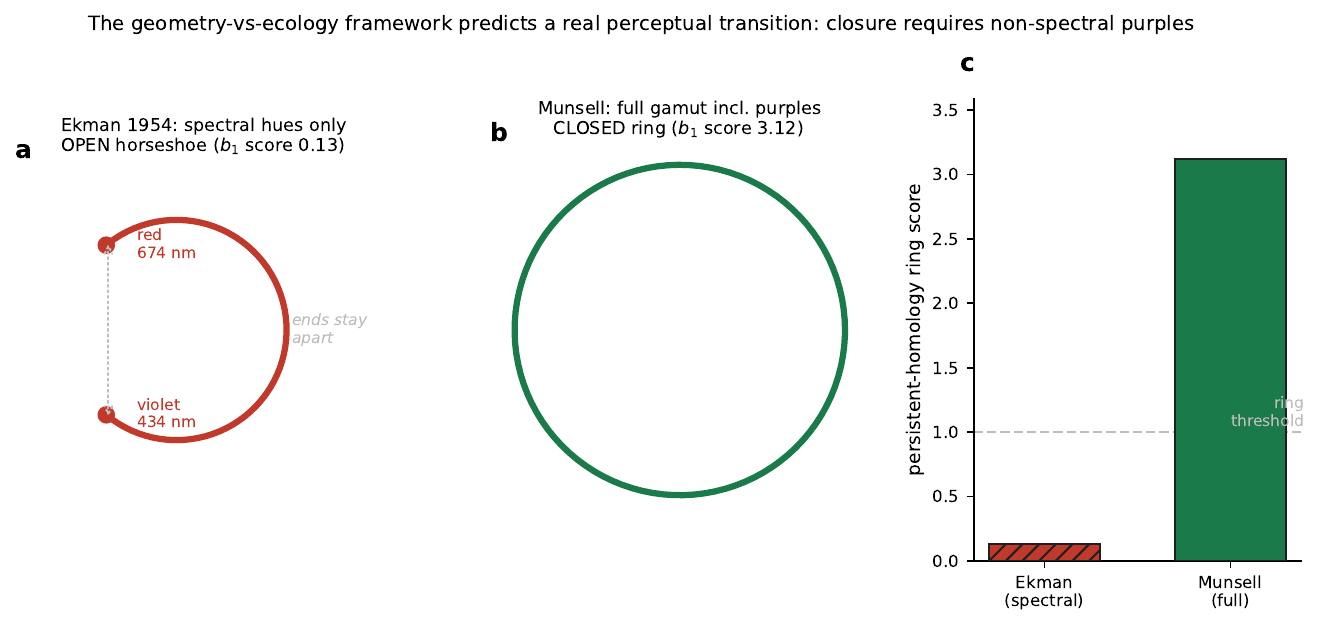}
\caption{\textbf{The geometry-versus-ecology framework predicts a real perceptual transition: closure requires non-spectral purples.} \textbf{(a)} Persistent homology on the \citet{ekman1954dimensions} similarity matrix of spectral hues yields an open horseshoe (ring score $0.13$): the red and violet spectral extremes are not adjacent. \textbf{(b)} The same analysis on the full-gamut Munsell renotation, including the non-spectral purples, yields a closed ring (ring score $3.12$). \textbf{(c)} The two ring scores against the persistence detection threshold; only the full gamut closes.}
\label{fig:perceptual}
\end{figure}

Finally, we asked whether the same topology machinery, applied to human perceptual data, recovers a known phenomenon. Human hue similarity is famously circular, but only when the full gamut is included. We ran our persistence pipeline on two classic datasets: the \citet{ekman1954dimensions} similarity matrix of fourteen \emph{spectral} hues, and the Munsell renotation covering the \emph{full} perceptual gamut including the non-spectral purples. The distinction is recovered directly from the data (Figure~\ref{fig:perceptual}): spectral-only similarity does not close into a ring (persistence ring score $0.13$; an open horseshoe whose red and violet ends stay far apart), whereas the full gamut does ($3.12$; a closed ring). We present this as a consistency check rather than a novel prediction: closure of the trichromat boundary is, at bottom, a geometric property of the bounded chromaticity region (in the sense of \citet{lee2024tetrachromatic}), and our modest contribution is that persistent homology recovers the open-versus-closed distinction directly from two classic perceptual datasets, tying the receptor-geometry framework to human color appearance, exactly as anticipated in Shepard's discussion of color as an internalized regularity~\citep{shepard1994perceptual}.

\section{Discussion}\label{sec:discussion}

We set out to ask two questions that had only ever been asked apart: what color geometry does a set of receptors afford, and how much of it does the natural world supply? Measuring both with persistent homology on real cone fundamentals, we find that the first grows with receptor number, tracing a hue ring, then a sphere, then a glome, exactly the \Snm{} ladder, and with a measured intrinsic dimension that tracks $n-2$, while the second stays nearly flat, so that the fraction of the available geometry the world actually fills declines steadily as receptors are added (slope $-0.056$ per receptor across 25 species, $p<10^{-4}$). A controlled decomposition locates the cause: the decline is driven by the fixed low spectral rank of the natural world, not by receptor count as such, and it disappears in a spectrally rich world even at high $n$. Its endpoint is the mantis shrimp, whose twelve receptors afford a topological ten-sphere but whose ecological color manifold is effectively two-dimensional.

The result reframes the dimensionality of color vision as a joint property of the eye and its environment rather than of the eye alone. Counting opsin classes measures what an animal \emph{could} encode; it systematically overstates what its world \emph{presents}. This does not contradict the analytic geometry of \citet{lee2024tetrachromatic} and \citet{logvinenko2025objectcolor}: we confirm and extend it, but it shows that the geometry is a ceiling, not a description. Nor does it contradict the long line of work showing natural color to be low-dimensional~\citep{maloney1986linearmodels,nascimento2005basisfunctions,chiao2000colorsignals,ruderman1998coneresponses}; rather it indexes that low dimensionality by receptor number and gives it a topological, comparative form, generalizing the within-human diminishing returns that \citet{foster2023colordeficient} report for red--green deficiency to the whole receptor series.

The most important relationship to prior work is with \citet{sabbah2013fourdimensional}, who argued that the spectral complexity of shallow-water light selected \emph{for} four-dimensional cone vision in fish. Their conclusion runs opposite to ours in direction, with complexity favoring more receptors versus the ecological manifold underfilling the receptors it has, but the two are not in conflict. Receptors may be selected for reasons the chromatic-manifold dimension does not capture: spatial and temporal signal detection, discrimination of specific ecologically loaded spectra, or communication signals~\citep{osorio2008review,kelber2010spectral,stoddard2008tetrahedral}, none of which require the full chromatic manifold to be uniformly populated. Our claim is narrow and geometric: whatever selected the receptors, the resulting high-dimensional hue manifold is not filled by natural spectra, and less so the more receptors there are. The mantis shrimp, whose receptors are widely thought to support rapid recognition rather than fine discrimination~\citep{thoen2014mantis,streets2022colour,wang2025opponency}, is precisely the case where this divergence is starkest.

Several limitations bound the claim and should be stated plainly. All publicly available hyperspectral scene databases are visible-only ($\geq 400$~nm); the ecological manifold for species with ultraviolet receptors (birds, insects, the mantis) is therefore computed over the visible range only, with the ultraviolet channels integrating over a spectral region the scene data do not resolve. We bounded this directly by recomputing the collapse on each ultraviolet species' \emph{visible-only} receptor complement (dropping every receptor peaking below $400$~nm; Figure~\ref{fig:expanded}d). Doing so \emph{raises} the collapse in every case: the blue tit from $0.67$ to $0.77$, goldfish $0.65\to0.84$, trout $0.62\to0.88$, butterfly $0.63\to0.71$, mantis $0.54\to0.70$, showing that part of the measured collapse in ultraviolet species is contributed by their ultraviolet channels sampling an unresolved region, so true ultraviolet scene statistics would make the collapse shallower. Crucially, the decline with receptor number \emph{survives} on the visible-only receptor complement, so the direction of our central result is not an artifact of ultraviolet truncation; only its magnitude for ultraviolet species is inflated, and we bound that inflation here rather than leaving it unquantified. For the mantis we report both the pigment-template receptor set (at measured $\lambda_{\max}$ values) and the measured stomatopod sensitivities digitized from \citet{thoen2014mantis}, which agree ($0.48$ vs $0.45$); the digitized curves cover eleven of the twelve channels (no data were recorded for the third R8 cell) and inherit the smoothing of the published figure. The other template species (goldfish, butterfly) were checked for robustness to $\lambda_{\max}$ placement; the remaining species use fully measured curves. High-dimensional persistent homology for $n\geq5$ relies on landmark witness complexes, and effective dimension rather than full Betti computation carries the argument beyond $n=5$. And the dissociation we observe between receptor count and behavioral discrimination acuity, found across the six species with comparable published wavelength-discrimination thresholds where more receptors do not buy finer discrimination and the twelve-receptor mantis is the coarsest ($\sim15$~nm), is a ranking/existence argument over a handful of published values, not a powered regression, and we do not advance a quantitative ecological-dimension-to-acuity law (the species are few and the ecological effective-dimension range is narrow). None of these caveats affects the core, repeatedly controlled finding: the collapse of the ecological manifold below the geometric bound, deepening with receptor number.

The framework also connects to perception. That closure of the human hue circle requires the non-spectral purples, with spectral hues alone forming an open arc (Figure~\ref{fig:perceptual}), follows from the geometry of the object-color-solid boundary and matches the classic view of color structure as an internalized regularity of the world~\citep{shepard1994perceptual,ekman1954dimensions}. We are deliberately modest about what this arm shows: the closure of the trichromat boundary is, at bottom, the convex closure of a bounded two-dimensional chromaticity region, geometric in the sense of \citet{lee2024tetrachromatic} rather than a novel statistical prediction, and our contribution here is simply that persistent homology recovers the open-versus-closed distinction directly from two classic perceptual datasets, tying the receptor-geometry framework to human color appearance. The ecological analysis bears on the \emph{filling} of the manifold, not on the existence of its closure. Where a future account of human color is likely to depart from pure geometry is in the metric rather than the topology: the strong anisotropy of human hue discrimination reflects the frequency of hues in the environment~\citep{hedjar2026environmental} and the placement of unique hues is not simply predicted by cone ratios~\citep{webster2000uniquehues,wuerger2005coneinputs}. Our contribution is to the topology and its dependence on receptor number; the metric that the ecological distribution imposes within the manifold is a natural next question. More broadly, the approach of reconstructing the topology of a stimulus manifold from data with persistent homology extends to color the program that has mapped the toroidal and ring topologies of neural population codes~\citep{gardner2022toroidal,singh2008topological,carlsson2008naturalimages}, and suggests that ``how many dimensions'' is, for any sensory system, a question that must be answered about the world and the receptor together.

\section{Materials and Methods}\label{sec:methods}

\paragraph{Receptors.} We used measured spectral sensitivities wherever public curves exist: the Stockman--Sharpe $2^\circ$ cone fundamentals for humans, and the measured visual-system sensitivities for dog, honeybee, blue tit, starling and peafowl from the \texttt{pavo} dataset. For species without a complete public measured set (goldfish, butterfly, mantis shrimp) we used Govardovskii visual-pigment templates~\citep{govardovskii2000template} placed at literature $\lambda_{\max}$ values, noted per species. All sensitivities were resampled to a common $1$-nm grid ($300$--$700$~nm) and normalized.

\paragraph{Why chromaticity, and why its boundary.} Receptor excitations live in an $n$-dimensional space, but one of those dimensions encodes luminance (overall intensity), which varies independently of surface color and is not a hue dimension in the ecologically relevant sense. We therefore work in \emph{chromaticity}, the $(n-1)$-dimensional space obtained by projecting each response vector onto the unit simplex (or equivalently, normalizing by the summed excitation). This is the standard reduction used in color science to isolate chromatic from achromatic variation~\citep{schrodinger1920pigmente,macadam1935optimal}. Within that $(n-1)$-dimensional chromaticity space the set of physically realizable colors forms a convex body (the chromaticity projection of the object-color solid), and it is the \emph{boundary} of this body, the locus of maximally saturated, spectrally extreme colors, that carries the topological signature \Snm{}. The boundary is the natural object: interior points correspond to mixtures and desaturated stimuli, whose geometry is controlled by the boundary; the topological invariant \Snm{} characterizes the hue circle/sphere/glome structure and is $n-2$ dimensional because we have already removed both the luminance direction (going from $n$ to $n-1$) and the interior of the chromaticity body (going from a filled $(n-1)$-dimensional solid to its $(n-2)$-dimensional surface). For a trichromat: 3 receptors $\to$ 2-D chromaticity $\to$ 1-D boundary = the hue ring $S^1$. For a tetrachromat: 4 $\to$ 3-D $\to$ 2-D surface = the hue sphere $S^2$. This chain is the geometric content of the \Snm{} claim; we use it consistently throughout.

\paragraph{Geometric manifold.} The geometric color manifold is the boundary of the object-color (Schrödinger--MacAdam optimal-color) solid~\citep{schrodinger1920pigmente,macadam1935optimal}: the convex set of chromaticities realizable by physical reflectances (values in $[0,1]$) under a fixed illuminant, seen through the species' receptors. We sampled optimal-color reflectances (single- and double-transition step functions plus random band reflectances), projected them to chromaticity, and extracted the boundary of the resulting convex body. Because the \Snm{} manifold is the boundary and not the filled solid, boundary extraction is essential and must be uniform: for two-dimensional chromaticity we densified the convex-hull polygon; for higher dimensions we sampled hull facets with probability proportional to facet area.

\paragraph{Ecological manifold.} Natural reflectance spectra were drawn from three independent hyperspectral databases, specifically Foster (Manchester natural scenes), Harvard (real-world indoor/outdoor), and CAVE (Columbia multispectral), converted to per-pixel reflectance and resampled to the common grid. These were multiplied by natural illuminants (Planckian, $4000$--$20{,}000$~K, $n=24$) and projected through each receptor set to chromaticity, yielding $1.5$--$2\times10^5$ samples per species. A uniform synthetic reflectance set (smooth sigmoidal bases) served as the negative control.

\paragraph{Persistent homology.} We computed persistent homology with Ripser, using a greedy-landmark (witness) complex for chromatic dimension $\geq 3$ to keep computation tractable. The diagnostic homology dimension is $n-2$. Betti numbers were called with a significance threshold calibrated against a matched Gaussian null (a feature counts only if its persistence exceeds $1.5\times$ the maximum null persistence and $15\%$ of the connected-component scale). The pipeline was validated on synthetic controls: circle ($b_1=1$), sphere ($b_2=1$), torus ($b_1=2,b_2=1$), filled disk and ball ($0$), and disk/ball boundaries recovering circle/sphere, passing all seven.

\paragraph{Effective dimension and collapse.} Effective dimension is the participation ratio of the chromaticity covariance, $(\sum_i \lambda_i)^2/\sum_i \lambda_i^2$ for eigenvalues $\lambda_i$. The collapse is the ratio of ecological to geometric effective dimension. The primary uncertainty we report is the standard deviation across the three hyperspectral databases (the ecologically meaningful variation); the seed-to-seed bootstrap over 8 seeds is far smaller and is used only where a single database is analyzed. The trend of collapse with receptor number is summarized by an ordinary-least-squares slope with a bootstrap 95\% confidence interval, in preference to a monotonicity claim. For the reconciliation with \citet{chiao2000colorsignals} we additionally report the number of principal components required to reach 95\% and 97.7\% of variance and the variance in the top three components.

\paragraph{Comparative series and dimension measures.} The comparative series covers \textbf{25 species} with published peak sensitivities: eight dichromats (dog, cat, horse, ferret, rabbit, pig, cattle, and the ultraviolet mouse), six trichromats (human, macaque, squirrel monkey, honeybee, bumblebee, jumping spider), nine tetrachromats (blue tit, starling, peafowl, chicken, zebra finch, pigeon, goldfish, rainbow trout, turtle), the butterfly pentachromat, and the mantis. Peak wavelengths are from standard comparative-vision sources (Jacobs; Neitz; Carroll; Hart 2001; Bowmaker; Kelber 2003; Peitsch 1992; Arikawa; Cronin/Marshall/Thoen). For each species the collapse was computed as the mean across the three hyperspectral databases and expressed under three distinct dimensionality functionals of the chromaticity covariance spectrum: the participation ratio (primary), a Shannon/entropy effective dimension $\exp(-\sum_i p_i\ln p_i)$ with $p_i=\lambda_i/\sum_j\lambda_j$, and the (continuously interpolated) number of principal components reaching 90\% of variance. The collapse--$n$ relationship was summarized by an OLS slope with a species-bootstrap 95\% confidence interval (resampling species) and by Spearman $\rho$ with a permutation $p$-value, and re-run on the measured-cones-only, $n\geq3$, and $n\leq5$ subsets.

\paragraph{Joint template perturbation.} To test whether the comparative trend depends on the literature peak wavelengths, we perturbed the peak of every template receptor in every species simultaneously by $\pm10$~nm and each template width by $\pm15\%$, and recomputed the whole collapse--$n$ relationship for each of 21 joint configurations, reporting the distribution of the slope and $\rho$.

\paragraph{Habitat prediction and ultraviolet bound.} Species were tagged by light habitat (aquatic, aerial, terrestrial) and, where a defensible value exists, by minimum wavelength-discrimination threshold. The prediction that spectrally narrower habitats collapse more at matched receptor number was tested by a one-sided Mann--Whitney $U$ comparing aquatic versus aerial/terrestrial tetrachromats ($n=4$) and ultraviolet versus non-ultraviolet trichromats ($n=3$). Because every hyperspectral database is visible-only ($\geq400$~nm), we bounded the ultraviolet confound by recomputing each ultraviolet species' collapse on its visible-only receptor complement (dropping every receptor peaking below $400$~nm) and confirming the decline with $n$ persists.

\paragraph{Intrinsic dimension.} To place the geometric and ecological manifolds on a common measuring stick, we estimated the intrinsic dimension of each with the maximum-likelihood estimator of \citet{levina2004maximum} (averaged over neighbourhood sizes $k=10$--$20$) and the two-nearest-neighbour estimator of \citet{facco2017estimating}. Both were validated to recover $1,2,3,4$ on synthetic $S^1,\dots,S^4$ and on filled balls (the MLE to within $10$--$15\%$; the two-NN estimator carries a known upward bias above $\approx3$D, so the MLE is used for the reported values). For the geometric manifold we estimated the intrinsic dimension of the extracted boundary point cloud; both estimators saturate near $5$D and the convex-hull boundary is degenerate in high ambient dimension, so intrinsic dimension is reported only for $n\leq5$ and the mantis argument rests on effective dimension.

\paragraph{World-rank decomposition.} To separate the world's spectral rank from receptor number as the cause of the collapse, we generated synthetic reflectance ensembles spanning exactly $r$ smooth spectral degrees of freedom (the first $r$ half-cosine basis functions on the grid, random coefficients, squashed to physical reflectance in $[0,1]$), for $r\in\{1,2,3,4,6,8\}$, and crossed them with Govardovskii receptor sets of size $n\in\{3,4,5,12\}$ at both evenly-spaced and clustered $\lambda_{\max}$, recomputing the collapse for each cell (3 seeds).

\paragraph{Mantis receptor robustness.} We recomputed the mantis collapse over fifteen receptor configurations: the baseline twelve-channel set, twelve independent jitters of all peak wavelengths by $\pm10$~nm, and two template-width perturbations ($\pm15\%$), reporting the mean and range.

\paragraph{Measured mantis sensitivities.} As a template-free check, we digitized the eleven intracellularly-recorded spectral sensitivity curves of \emph{Haptosquilla trispinosa} from Figure~1A of \citet{thoen2014mantis} by axis-calibrated colour segmentation (each curve assigned by nearest legend colour; the plot was rendered at $12\times$ resolution and the two axes calibrated from their tick marks), then interpolated gaps, lightly smoothed, and unit-area-normalized each curve exactly as for the other species. The recovered peak wavelengths span $317$--$604$~nm and reproduce the published ordering; the collapse was then computed with the identical pipeline. The third R8 channel was omitted because no sensitivity was recorded for it in the original.

\paragraph{Discrimination dissociation.} For the six species with a defensible published minimum wavelength-discrimination threshold ($\Delta\lambda$), we correlated receptor count, geometric effective dimension, and ecological effective dimension with discrimination acuity ($1/\Delta\lambda$) by Spearman $\rho$ with a bootstrap 95\% confidence interval and a permutation $p$-value, treating the result as a ranking argument rather than a powered regression.

\paragraph{Receptor noise.} Behavioral discriminability was modeled with the Vorobyev--Osorio receptor-noise-limited transform~\citep{vorobyev1998receptor} at Weber fractions $\{0.02, 0.05, 0.1\}$, applied in log-quantal-catch space before recomputing effective dimension.

\paragraph{Perceptual validation.} The same persistence pipeline was applied to the \citet{ekman1954dimensions} $14\times14$ spectral-hue similarity matrix (as a distance matrix) and to the Munsell renotation full-gamut chromaticities, reporting the ring score (longest $H_1$ bar relative to the connected-component scale).

\paragraph{Compute and reproducibility.} Analyses ran   with fixed random seeds. Code and the figure source data are openly available at \url{https://github.com/rostami-m/Color_Vision_Dim}.

\paragraph{Use of generative artificial intelligence} 

During the preparation of this manuscript, the author used a large language model for language editing, grammar correction, sentence restructuring,  improving the clarity and flow of the scientific narrative with light conceptual structuring.  All AI-generated   suggestions were critically reviewed, fact-checked against the underlying empirical results, and approved by the author.

\section*{Acknowledgements}
The author thanks the maintainers of the Foster, Harvard and CAVE hyperspectral databases, the \texttt{pavo} project, and the CVRL color database for making their data openly available.

\section*{Additional information}

\subsection*{Competing interests}
The author declares that no competing interests exist.

\subsection*{Funding}
This work received no specific external funding.

\section*{Data availability}
This is a computational study using only previously published, openly available datasets; no new empirical data were generated. All code required to reproduce the manifolds, topology, analyses, and figures is openly available under the MIT License (archived at Zenodo). Spectral sensitivities are from the Stockman--Sharpe/CVRL database and the \texttt{pavo} package; hyperspectral reflectances are from the Foster, Harvard and CAVE databases; perceptual data are the Ekman 1954 similarity matrix and the Munsell renotation. All are publicly downloadable and fetched automatically by the released code.

\bibliography{references}

\end{document}